# Ultrafast solvation dynamics at internal site of staphylococcal nuclease investigated by site-directed mutagenesis

Gao Guang-yu(高光宇)[1], Li Yu(李渝)[1], Wang Wei(王伟)[1], Wang Shu-feng(王树峰)[1,*], Dongping Zhong [2], Gong Qi-huang(龚旗煌)[1,3]

[1] *Institute of Modern Optics & State Key Laboratory for Artificial Microstructure and Mesoscopic Physics, School of Physics, Peking University, Beijing 100871, China*

[2] *Departments of Physics, Chemistry and Biochemistry, Programs of Biophysics, Chemical Physics and Biochemistry, The Ohio State University, Columbus, OH 43210, USA*

[3] *Collaborative Innovation Center of Quantum Matter, Beijing, China*

Solvation is essential for protein activities. To study internal solvation of protein, site-directed mutagenesis is applied. Intrinsic fluorescent probe, tryptophan, is inserted into desired position inside protein molecule for ultrafast spectroscopic study. Here we review this unique method for protein dynamics researches. We introduce the frontiers of protein solvation, site-directed mutagenesis, protein stability and characteristics, and the spectroscopic methods. Then we present time-resolved spectroscopic dynamics of solvation dynamics inside caves of active sites. The studies are carried out on a globular protein, staphylococcal nuclease. The solvation at internal sites of the caves indicate clear characteristics of local environment. These solvation behaviors correlated to the enzyme activity directly.

## 1. Introduction

Proteins and water are essential to life on earth, since they are the key factors in all biological activities. The proteins function as structural elements, catalysis of biological reaction, carriers for ions and molecules, immunization antibodies, and luminance emitters et. al. The complication of their structures and environment dependency are of great interest to the understanding of their biological functions.[1] Among all the environmental interaction, the solvation is the one of the most important issues, since biological processes happen under the water environment. The water molecules couple to the protein structures and motions so that the proteins can present their biological activities.

The water in biological system has various functions affecting or participating the biological processes. Ordered water molecules accompanying the protein framework help the folding and stabilizing the protein structure.[2] In membrane system, the water helps the element transportation.[3] In reaction of hydrolyzing DNA and RNA by staphylococcal nuclease (SNase), the water molecule is electron donor for nucleophilic attack.[4] It can also be a proton donor for biological reactions.[5] For all these studies, solvation dynamics is one of the most important aspects to understand the protein-solvated water interaction and corresponding dynamics.[6,7] The meaning of protein solvation dynamics is the localized response of polar solvent molecules to the protein with static charge distribution at non-equilibrium state. Through the adjustment of water molecule motions, including translation, vibration, rotation, exchange et. al., the non-equilibrium charge distribution becomes equilibrium by coupled motion of protein and water molecules. This is also called hydration dynamics. By hydrogen bond, charge-dipole, or dipole-dipole interaction, the water molecules bond to the protein, forming solvated water at protein surface, cavities, and internal sites. Because of these strong interactions, their dynamics strongly relate to the local protein framework motion and residue characteristics, which is dramatically different to the bulk water. Finally, the water participates the protein function or drives the protein folding to form stable structures.

---

* Corresponding author. E-mail: wangsf@pku.edu.cn

The protein dynamics is from the ultrafast time scale of chemical bond vibration to seconds for protein folding, while the protein solvation dynamics is from sub-picosecond to nanosecond. Ultrafast fluorescent spectroscopic techniques are fundamental for this study. Intrinsic or external fluorescent probes are applied to obtain the ultrafast response of the protein towards femtosecond disturbing.[8] The internal conversion of these probes is usually less than 100fs, while their fluorescent lifetimes extend to nanoseconds. Therefore, these probes provide proper observation window for studying protein solvation dynamics.[9-11] The intrinsic probe usually used is tryptophan (Trp, or, W), which has red-most emission band among twenty amino acids. By mutagenesis, the Trp can be inserted into desired positions of protein framework, in order to study site-specified protein dynamics. This is more versatile than applying external probes.

The solvated water can be roughly classified to surface water and internal water. They have dramatically different dynamical behaviors. The surface water is also called bound water, since it has direct interaction to the surface residues and form hydration layer. Comparing to ~100fs and few picoseconds processes in bulk water,[10,12] the bound water also has two processes of ~1ps and ~100ps,[13,14] which is correspondent to two types of motions, vibrational and rotational motions of water molecules in hydration layer, and translational and exchange motions between bound water and bulk water, together with the coupled adjust motion of protein framework.

The most systematical study on surface water hydration dynamics was carried out by Zhong's group.[15] They applied ultrafast fluorescent up-conversion technique to investigate the surface of protein apo-myoglobin, site by site. They found that the hydration dynamics are clearly different towards the depths of Trp probe at the protein surface. It is also affected by local property of protein for each site. Two emission peak shift (Stocks shift) are resolved as $\Delta E_1$ and $\Delta E_2$. The $\Delta E_1$ increases monotonically when the probes become closer to protein surface. This means the probe is gradually sensing the whole hydration layer. When the probe becomes closer to the surface, it is exposed to stronger polar environment. $\Delta E_2$ does not show this monotonic increment. Instead, it increases to a fixed value when gradually exposed to the surface. This limited and saturable shifting is due to the limited interface water molecules bond to protein surface. When the probe is buried deeply inside protein, it has non-polar local environment with shortest emission wavelength less than 330nm. At intermediate depth, it shifts to 330-340nm. When the probe become close to the surface, it has emission peak longer than 340nm, but the dynamical emission shift keeps ~600 cm$^{-1}$ at this depth. Besides the two shifts, the two time scales for the shift are also informative. They do not show monotonic relation to the amplitude of shift. For the first time scale of few picosecond, $\tau_1$, it represents the local motion of water molecules, including the vibration and rotation. Some sites with faster $\tau_1$ (1-2ps) are surrounded by the flexible loop structure, where the water molecules are quite mobile. Slower $\tau_1$ happens at sites with intensely charged/polar amino acid residues, where the water molecules are bounded and cannot move so freely. For overall trend, $\tau_1$ becomes faster when the emission spectra shift to the red. The slower time scale, $\tau_2$, which is ~100ps, is the re-arrangement of the water molecular network in hydration layer. Similar to $\tau_1$, the time scale of $\tau_2$ is sensitive to the rigidity of protein framework. In high charge density sites, the $\tau_2$ is large. This study gives an overall dynamical picture for hydration water at outer surface of protein.

The water buried inside protein molecules is also important for structural integrity and functionalities. These water molecules take various roles. In polar cavity, the water adjust the protein folding and their stability, even the catalysis of protein.[16] It has been proved, both in experiments and molecular dynamics simulation, that the water molecules exist inside protein, even inside the hydrophobic cores.[17-21] Several techniques are applied on this issue, including nuclear magnetic resonance (NMR), molecular dynamics simulation (MD), and time-resolved emission spectroscopy (TRES) et. al. Comparing to the solvation dynamics study at protein surface, the dynamics of internal water are not well understood at the moment. Inside caves with entrance at the protein surface, there are limited water molecules for hydration.[4,22,23] Some studies conclude their dynamic behaviors differ to the surface greatly, which depend on the polarity of the cavity. The NMR studies show their solvation is around nanosecond.[17,18] It is observed that the exchange of internal water is very slow, up to nano- or microsecond scale.[24,25] However, others believe that the residence time should be in picosecond time scale.[26] The ultrafast spectroscopic study with fluorescence probe inside protein indicated no slow process, but similar behavior to surface water.[15] Qin found that water motion in the binding pocket of polymerase-Dpo4 is a biphasic behavior of a few picoseconds to tens of picoseconds similar to that of surface water.[26] Chang claimed that the local motion of water in different pockets in photolysis is a triphasic behavior, two fast motions from a few picoseconds to tens of picoseconds, plus a sub-nanosecond motion.[6] The two fast ones are similar to those of surface water, while the slowest one may relate to the strong bonding to the cavity. However, this last process may not result from solvation, since the water molecules inside cavity are very limited. In the study of protein-precipitate system, or protein-DNA system, the probe at protein surface is enclosed at a small space similar to the cavity. In this situation, long decay component was not found.[27] These results show unique dynamic properties of the coupling between internal water motion and protein folding or enzyme activity. Direct optical observations on solvation dynamics are still limited, especially in the interior of protein. There are only a few studies on the local dynamics of protein active regions, as well as their hydrophobic core.

2. SNase, tryptophan, and mutagenesis

SNase is a widely studied model protein for protein folding,[28-31] stability,[32,33] dynamics,[34,35] and catalysis activity.[36] SNase, which comes from staphylococcus aureus, is a small globular protein with single chain and single domain. Its molecular weight is 16800kDa, with 149 amino acid including one Trp residue at site 140. It has two subdomains, N-terminal and C-terminal subdomains, which contain five β-strands and three α-helix, respectively. It non-specifically hydrolyze single or double strand DNA and RNA, producing nucleoside 3'-phosphates and 3'-phosphooligonucleotide. Its active domain contains two pockets, nucleotide binding pocket and $Ca^{2+}$ binding pocket.

There are three native aromatic amino acids with fluorescence, phenylalanine (Phe), tyrosine (Tyr), and tryptophan (Trp). The Trp has the largest extinction coefficient, moderate fluorescent quantum yield, and red-most absorption.[37] It is very sensitive to local polarity variation. The conformation change, substrate bonding, and protein folding et. al. can affect its emission spectra. Its indole ring is also a good electron donor, which can easily be quenched by neighboring amino acid. Therefore, the Trp is suitable as a native fluorescent probe for studies on protein structure and function.

The absorption band of Trp locate at ~280nm, where overlap that of tyrosine. Pumping at 295nm can avoid the overlap and excite Trp only. Deeply buried in hydrophobic core, its emission has the shorted wavelength, e.g., Trp48 in azurin has emission at 308nm.[38] At protein surface where it exposes to water, the emission maxima are close to 350nm. Trp has two singlet excited states, $^1L_a$ and $^1L_b$, whose dipole moments are nearly perpendicular to each other. In non-polar solvent, $^1L_b$ is the lower state, from which the main fluorescence comes. In water solution, the $^1L_a$ is lower, and almost all fluorescence is from $^1L_a$. The fluorescence decay may be affected by the neighboring amino acid et. al. Therefore, multiple decaying processes can be observed, which should be identified in solvation dynamics studies.[39]

To insert Trp into desired sites of SNase, protein mutagenesis is applied. The wild type (WT) SNase (PDB: 1SNO) has a single native Trp at position 140. Firstly we replaced this Trp in WT by histidine (H) to generate a single point mutant, W140H. The W140H has good productivity and stability as a template for further modification. The target SNase has a single Trp residue inside the protein, which replaces one original amino acid residue of desired site to generate dual-point mutants. The designing of Trp site has to avoid the quenching effect from the amino acid with carboxyl group, such as glutamic acid. Then we designed twenty-five mutated SNase with a single Trp in each. All the sequences of the mutants were confirmed before further purification. The mutants were expressed in Escherichia Coli BL-21(DE3) and purified by the procedures described in Ref. [14]. The protein concentrations were determined by the absorbance at 280 nm.[40]

For all these double point mutants, they have clear absorption structure at 290nm, while for template, W140H, whose Trp is missing, there is no strong absorption at this wavelength. The W140H also shows no emission peak as all other mutants, excited at 295nm.

3. **Spectroscopic method and solvation correlation function**

The solvation process can be observed by inspecting the emission spectra from probes. The ultrafast time-resolved emission spectra (TRES) can be obtained by fluorescent up-conversion technique at multiple wavelengths with time resolution up to ~100 femtoseconds, or even shorter. By streak camera system, since both the temporal and spectral signals can be record simultaneously, TRES can be recorded directly. However, its time resolution is limited to picosecond timescale.

The transient fluorescent decay of Trp contains two typical processes. The first one is from solvent relaxation, which has two decays, a picosecond one and a hundred picosecond one. The second is the native relaxation of Trp, so-called lifetime decay, which may also contain two processes both at nanosecond time scale.

$$I_\lambda(t) = I_\lambda^{solv} + I_\lambda^{popul} = \sum_i amp_i \exp\left(-\frac{t}{\tau_i}\right) + \sum_j amp_j \exp\left(-\frac{t}{\tau_j}\right) \quad (1)$$

The *i* is solvation process while *j* is the contribution of lifetime. In solvation process, at blue side, the $amp_i$ is positive, while at red side, it is negative, indicating raising of intensity. The lifetime components are always positive in this case. Through deconvolution, we obtain the peak shift for reconstructing TRES, *ν*(*t*). Then, to achieve time-resolved Stocks shift of solvation process only, we need to remove the shift from native decay related to lifetime.

$$I_\lambda^{popul}(\lambda, t) = \frac{I_\lambda^{popul}(t)I_\lambda^{SS}}{\sum_i amp_i\tau_i + \sum_j amp_j\tau_j} \quad (2)$$

The $I_\lambda^{SS}$ is the steady state spectral intensity at *λ*. By this way, we obtain the TRES of native decay processes, *ν*<sub>*l*</sub>(*t*). Therefore, the solvation correlation function *C*(*t*) can be built as

$$C(t) = \frac{\nu(t) - \nu_l(t)}{\nu(0) - \nu_l(0)} \quad (3)$$

This correlation function can be fitted by dual-exponential function, which are correspondent to spectral shift, Δ*E*$_1$ and Δ*E*$_2$.

## 4. The steady state spectroscopy of SNase mutants and their structural integrity

The fluorescence of Trp are widely used to reflect the local environment through their emission maxima, λ$_{max}$. It is usually regarded that when λ$_{max}$ is shorter than 330nm, the Trp is in hydrophobic local condition. When it is between 330 and 338nm, the probe is partially exposed to the polar environment. When is longer than 338nm, it is thoroughly exposed to polar surroundings. As our mutants, though we targeted to the internal sites of SNase, the partially exposure of water environment can also be observed, through fluorescent emission spectra, as shown in Fig. 1.

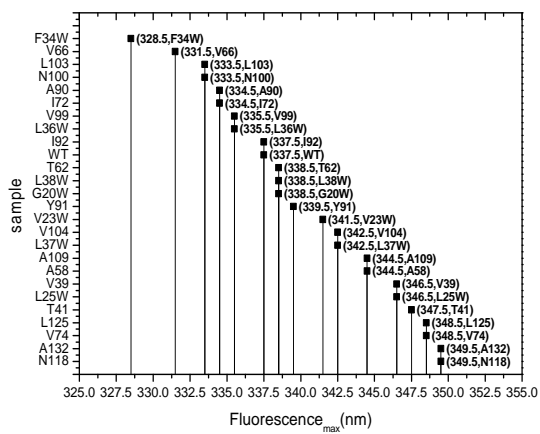

Figure 1. fluorescent emission maxima of Trp in SNase mutants.

The F34W and V66W present blue most emission spectrum at 328nm and 330nm respectively, indicating a strong hydrophobic local environment, while the others present gradually exposure to water. N118W has red most emission of 350nm, which means it is at a highly exposed position. The WT SNase has the emission at 337nm. The Trp in this WT is on a loop structure and partially exposed to water.

The Trp has hydrophobic residue, and is largest one among twenty amino acids. The mutation with Trp then may experience conformation variation.[33,41-43] According to the emission spectra, the Trp of L36W should be in deeply buried hydrophobic core in β-barrel, similar to F34W. It should show similar blue emission when it is in hydrophobic environment, i.e. less than 330nm. The study show that it is at 334nm, partially exposed to water, similar to WT. This case is also found for V23W, L25W, L37W, L38W, and V39W et. al. However, most mutant keeps their overall structures similar to WT. By checking the structural integration through various methods, we found that most of the mutants has loosen but similar structure to WT. The loosening is induced by the large residue of Trp which become more or less exposes to surrounding water. However, to our study on cavity hydration, the four selected mutants are in correct spatial location, which will be discussed later.

We carefully choose thirteen double point mutants to study their structural integrity. They are G20W, V23W, L25W, F34W, L36W, L37W, L38W, V39W, T41W, V66W, N100W, L103W, and N118W. By checking the circular dichroism (CD) spectra, enzyme activity, and surface hydrophobicity by external probe (1-anilino-8-naphthalene sulfonate, ANS), we determined that all the mutants are relatively loose to the WT, while their enzyme activity are kept at certain level. This means that the structural integrity are retained. Among the mutants, V23W, L25W, F34W, and L36W are somewhat more loosened than the others, which are in a status usually called molten globule, while the others are in between native folding and molten globule.

When these mutants bond to substrates, their structures can be partially recovered. A substrate analog, pdTp, is applied in the study. PdTp has stronger affinity to the pockets of active site towards DNA or RNA, so it is called inhibitor. By detecting ANS fluorescence, nearly all mutants, except V39W and L103W, present weakened hydrophobic environment after binding with pdTp, which means the interaction between proteins and substrate, pdTp (or DNA/RNA), drives proteins to fold into a more WT-like form. After binding to pdTp, F34W, V66W, and N118W has very similar surface hydrophobicity to WT. This is also proven by CD spectra.

5. **Internal solvation dynamics of SNase**

The study of internal solvation dynamics is fundamental to understand the dynamic mechanism of protein structure, recognition, and enzyme activity. The existence and localized motion of internal water maintain structural integrity of protein, and further optimize the local spatial configuration for interacting with ligand or substrate.[24,44,45] The internal water may also take part in the catalysis process, e.g., as a donor in nucleophilic attack for hydrolysis of DNA and RNA.[4] Therefore, investigating the solvation of the internal water is a way to understand the protein dynamics. In this section, we will discuss the detailed experiment and result on internal cavity dynamics of SNase. We select four mutants, F34W, L36W, L38W, and T41W, as shown in Figure 2.

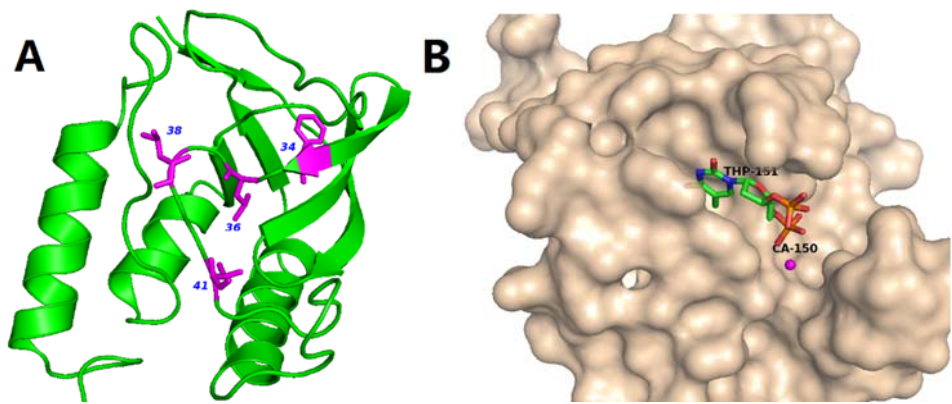

Figure 2 A: the X-ray crystal structure of SNase (PDB:1SNO). The positions substituted by Trp are denoted in purple sticks, at position 34, 36, 38, and 41. B: the active region of SNase is shown in surface mode, where the substrate analogue, pdTp, and ligand $Ca^{2+}$ are inserted to demonstrate the two binding pockets, shown in sticks.

Picosecond TRES measurements of SNase were conducted using an optically triggered streak camera system (C5410, Hamamatsu). All of the samples were excited at 295nm. Samples were filled into a 1-mm thick transparent rotating cell. The polarization of fluorescence from sample was selected by a Glan-Taylor prism at the magic angle (54.7°) with respect to the polarization of the pump beam. For fluorescence anisotropy measurements, the Glan-Taylor prism was rotated to be either parallel or perpendicular to the polarization axis of the pump beam to obtain the parallel ($I_\parallel$) and perpendicular ($I_\perp$) signals, respectively. These transients were used to construct the time-resolved anisotropy: $r(t) = (I_\parallel - I_\perp)/(I_\parallel + 2I_\perp)$. The energy of 295 nm pump beam was below 1 nJ per pulse. The width of instrument response function is around 6 ps and 60ps for two time windows (160 ps and 2200 ps), respectively. The final temporal resolution is around 3 ps upon deconvolution. The concentrations of all samples were 200 μM in dialysis buffer.

The lifetime measurements of Trp fluorescence were performed at room temperature with a spectrofluorometer (Edinburgh) at 370 nm. The time window and the peak counts are 100 ns and 2000ns, respectively. The full width at half maximum of the instrument response function is ~1.8 ns. The protein concentration is 50 μM. The lifetime were directly given through the software.

The steady-state fluorescence spectra were measured at room temperature. In this fluorescence study, the concentrations of the samples were 5 μM in dialysis buffer. The fluorescence were excited at 295 nm, while the emissions were recorded between 300-450 nm.

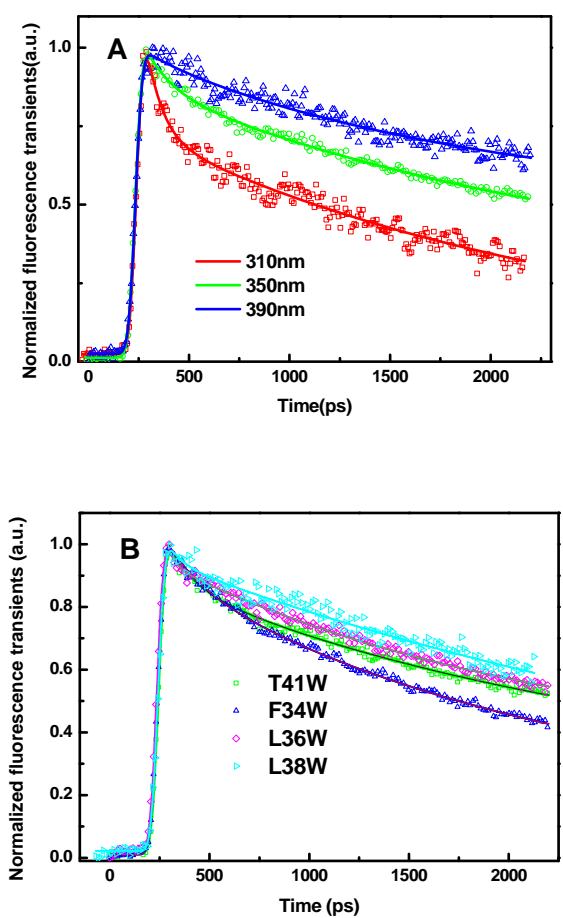

Figure 3 A: Normalized picosecond-resolved fluorescence transients of T41W for three emission wavelengths, 310 nm, 350 nm, and 390 nm. B: Normalized picosecond-resolved fluorescence transients of four mutants at 350 nm. The scattering dots are the original fluorescence transient data; the solid lines are the best fitting curves of the transients.

The normalized steady-state emission spectra of F34W, L36W, L38W, and T41W are 328 nm, 334 nm, 336 nm, and 346 nm, respectively, showing increasing local polarity for the four mutants. The local environments of the Trp probe in F34W, L36W, and L38W display very weak and comparable polarity, while T41W shows relatively stronger polar environment. The lifetime and quantum yield of WT and the mutants are listed in Table 1. All mutants display reduced quantum yields towards WT. Neglecting the contribution of ultrafast solvent relaxation on 1-100ps scale, all samples exhibit similar two-lifetime decays in nanoseconds, $\tau_{L1}$~1.5-3.2 ns and $\tau_{L2}$ ~ 5.0-6.8 ns. No ultrafast quenching process in sub-nanosecond exists.

Table 1: Fluorescence quantum yield of mutants relative to WT and lifetime

| Samples | $Q_{mutants}/Q_{WT}$ | $\tau_{L1}$ | $\tau_{L2}$ | $B_1$ | $B_2$ |
|---|---|---|---|---|---|
| WT | 1.0 | 2.8 | 5.6 | 0.18 | 0.82 |
| F34W | 0.296 | 2.0 | 5.0 | 0.69 | 0.31 |
| L36W | 0.44 | 2.0 | 5.6 | 0.57 | 0.43 |
| L38W | 0.45 | 2.5 | 6.6 | 0.58 | 0.42 |
| T41W | 0.36 | 2.3 | 6.8 | 0.63 | 0.37 |

Q: fluorescence quantum yield

$\tau_L$: lifetime in nanoseconds

B: the portion of lifetime, where $B_1+B_2=1$

Figure 3A shows the picosecond-resolved fluorescence transients of T41W for three typical wavelengths from the blue to red side of Trp emissions. All transients were taken within two time windows: 160 ps and 2200 ps. At shorter time window of 160ps, two ultrafast components, $\tau_1$ and $\tau_2$, in the range of several picoseconds and tens of picoseconds can be found besides the two intrinsic lifetime components $\tau_{L1}$ and $\tau_{L2}$. Similar to the reports,[9,10,27] it is a manifestation of the local solvation dynamics of neighboring water molecules. Systematic studies in these reports indicated that the decay in a few picoseconds represents the water-network local relaxation, while the decay of tens of picoseconds are the water-network rearrangements, coupled with local protein fluctuations, from a non-equilibrium configuration to an equilibrated state. The other three mutants have similar patterns of temporal behaviors. Their variation are shown in figure 3B.

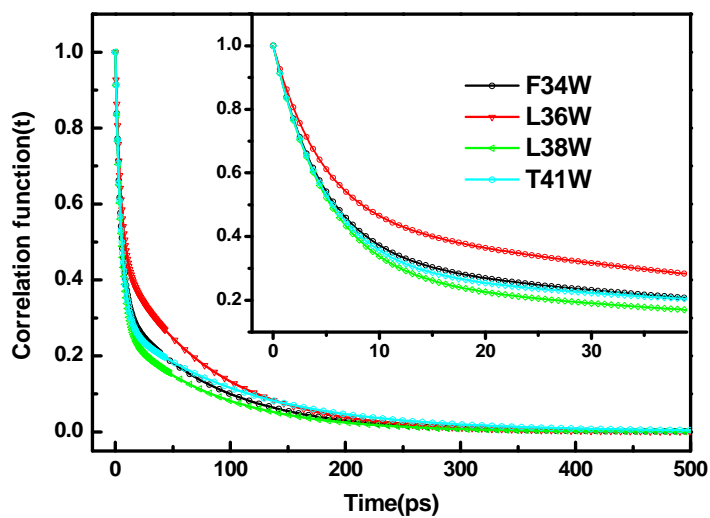

Figure 4 Solvation correlation functions of four mutants F34W, L36W, L38W and T41W are shown in symbol and line; inset: correlation functions within 40 ps.

Table 2: Results obtained from the hydration correlation functions $C(t)$ of four mutants of SNase

| Mutants | $\lambda_{max}$ | $\tau_1$ | $\tau_2$ | $C_1$ | $C_2$ | $E_1$ | $E_2$ |
|---|---|---|---|---|---|---|---|
| F34W | 328 nm | 4.6 | 76 | 0.42 | 0.58 | 78.9 | 108.9 |
| L36W | 334 nm | 5.0 | 77 | 0.30 | 0.70 | 69.4 | 162.0 |
| L38W | 336 nm | 5.2 | 90 | 0.63 | 0.37 | 196.7 | 115.5 |
| T41W | 346 nm | 9.8 | 104 | 0.42 | 0.58 | 258.3 | 356.7 |

All of the solvation correlation functions were fitted with $C(t)=C_1*exp(-t/\tau_1) + C_2*exp(-t/\tau_2)$, where $C_1+C_2=1$. The time constants are in picoseconds. $E_i$ is the dynamic stokes shift of $\tau_i$ component in cm$^{-1}$. $\lambda_{max}$ represents the emission maximum in steady-state.

We fitted the time-resolved fluorescence spectra with a log-normal function to deduce the dynamic stokes shifts $v_s(t)$. The lifetime-associated emission contribution $v_l(t)$ were obtained with three-exponential fitting on $v_s(t)$ and extrapolation to time zero. Figure 4 shows the solvation correlated dynamical Stokes shift of four mutants, $c(t)$, which was constructed following equation (1). All solvation correlation functions can be fitted by a dual-exponential decay, with two time constants $\tau_1$ (a few picoseconds) and $\tau_2$ (tens of picoseconds). The results are shown in Table 2. For all mutants, the two time constants of solvation dynamics are within 10ps and 110ps similar to the results probed by Trp at protein surface and the active regions,[14] reflecting that water in the binding pockets (or in the interior of cavity ) of SNase are not tightly trapped. The response from trapped water, as well as from the neighboring polar/charged residues, should be hundreds of picosecond to nanoseconds, as observed in reports, [6,46] which describe the slow collective motions.

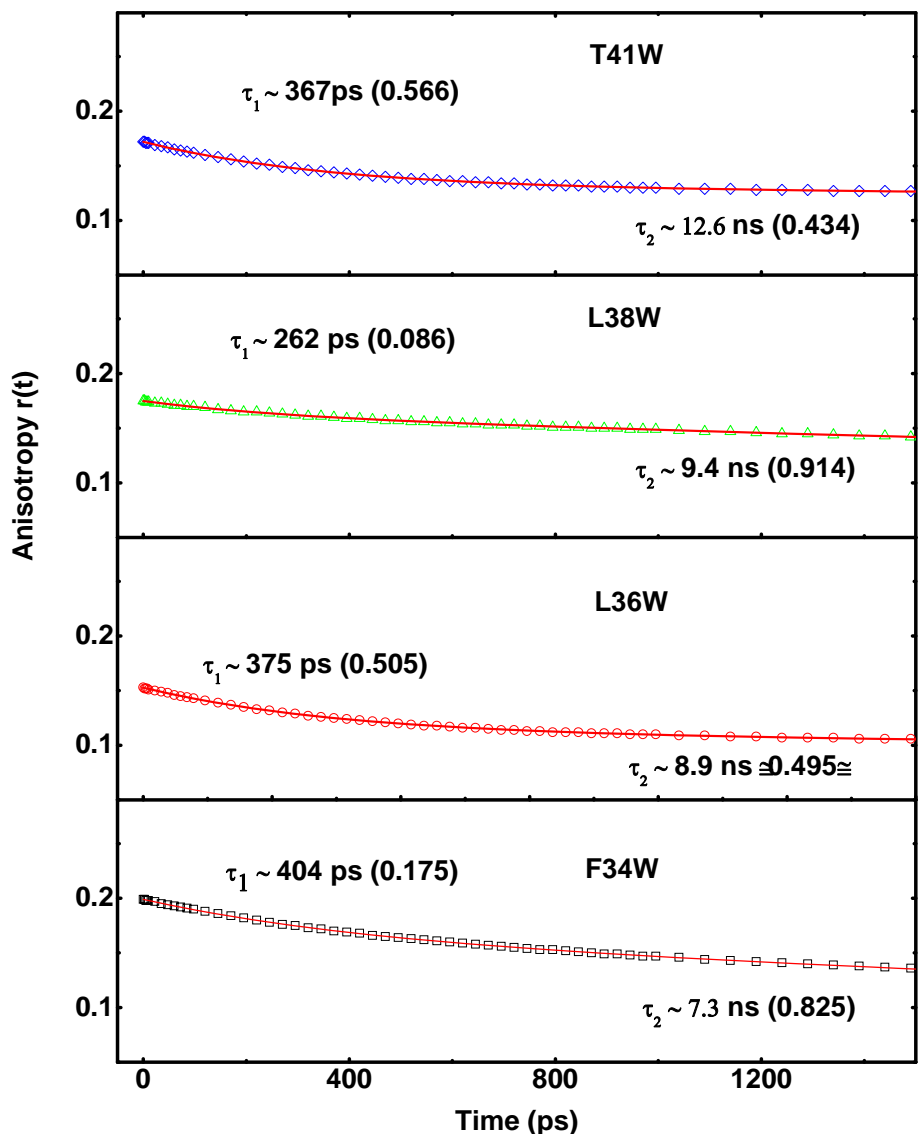

Figure 5: the picosecond-resolved anisotropy of Trp emission, together with fitting, using two-exponential function.

We also studied the picosecond-resolved rotational dynamics of Trp by probing the anisotropy, as shown in Figure 5. We observed that all anisotropy dynamics are biphasic behavior with a fast decay component in hundreds of picoseconds time scale and a slow component in several nanosecond time scale. The nanosecond dynamics from 7.3 ns to 12.6 ns represents the whole protein tumbling motion. The picosecond dynamics from 262 to 404 ps result from the local wobbling motions.[14] All of these findings suggest a relatively small local fluctuation. The local structure around Trp probe in the samples does not undergo large conformation changes in our measurements. The conclusion is indeed consistent with the results of the steady-state fluorescence emission maxima.

The time constants of solvation dynamics of the four mutants probed by Trp clearly reflect unique local dynamic properties in the interior of SNase. According to the crystal structure, Trp in F34W is deeply buried in the β-barrel with high hydrophobicity. It observes the fastest time constants of solvation dynamics accompanied with the least corresponding energies relaxation. This indicate that its local environment is in very weak polarity and low solvated so that content of energy relaxation which representing the extent of hydration is small. It is similar to the dynamic properties of the mutants W14, A15W and H12W in Ref. [15] Therefore, the dynamic results of F34W represent typical solvation features of internal hydrophobic region. However, Due to water existing in the hydrophobic cavity of protein,[47,48] we cannot thoroughly remove the contribution of solvation relaxation from protein surface and/or the nucleotide binding pocket of SNase. The probable internal water buried in the hydrophobic cavity of SNase [44] may also contribute to the solvation relaxation, but it is supposed to be very weak because of limited amount of water molecules. For L36W, the side-chain of Trp stretches into the β-barrel and is deeply buried below the nucleotide binding pocket. L36W shows quite similar time constants and energies relaxation of solvation dynamics to F34W. The results of F34W and L36W are consistent with their steady-state fluorescence data and similar to the results in Ref.[15], verifying that the inside of hydrophobic core should be of low solvated.

For the nucleotide binding pocket, Trp in L38W lies at the bottom of the pocket, where the Trp probe is mainly surrounded by several hydrophobic and polar residues, such as Leu-37, Thr-113, and Thr-115. L38W has slower time constants but bigger stokes shifts, $\tau_1$ (5.2 ps) and $\tau_2$ (90 ps) with $E_1$ 196.7 cm$^{-1}$ and $E_2$ 115.5 cm$^{-1}$, than F34W and L36W, as shown in Table 2. The first time constant $\tau_1$ of L38W is quite similar to that of polymerase mutant S244W in apo state,[27] revealing similar local dynamics and polarity of the nucleotide binding pocket for different enzyme. The solvation relaxation of L38W can be mainly assigned to the contribution from the hydrated water in the pocket. The slower time constants reveal moderate polarity at the bottom of the pocket, which confines the local motion of the cavity water. For T41W, the probe is located at the bottom of the Ca$^{2+}$ binding pocket. This pocket, surrounded by dense of charged and polar residues, including Asp-21, Asp-40, Thr-41 and Glu-43, is responsible for the hydrolysis of the phosphate backbone of DNA and RNA when Ca$^{2+}$ is in the pocket. The high polarity distribution in this region results in the slowest time constants of solvation relaxation $\tau_1$ (9.8 ps) and $\tau_2$ (104 ps) with $E_1$ 258.3 cm$^{-1}$ and $E_2$ 356.7 cm$^{-1}$ as shown in Table 2. The increases in solvation time constants and Stokes shifts relative to F34W, L36W and L38W represent more confined water was detected, including close probing of the containing water in the binding pocket and distant probing of surface water.

The distinct differences in time constants of solvation relaxation between nucleotide binding pocket and Ca$^{2+}$ binding pocket as well as the hydrophobic core in SNase clearly reveal characteristic dynamic features of different local structures. For a compact hydrophobic core (β-barrel), the biphasic behavior of solvation relaxation are similar to rigid regions, such as binding pockets, and flexible regions, such as loop patches. [16] For the nucleotide binding pocket,[4,49] the residues (Arg-35, Lys-84, Tyr-85 and Arg-87) and the water form hydrogen-bond networks to bind nucleotide at a perfect position. For L38W, the Trp probe detects the local dynamics of the pocket bottom, which is mainly surrounded by the two polar residues Tyr-113 and Tyr-115, and other nonpolar ones like Leu-37. (Tyr-113 and Tyr-115 in nucleotide binding pocket may have π-π interactions with the adenine part of nucleotide to further regulate its binding.) Thus this region is in moderate polarity and solvated. Local water molecules are not confined tightly compared to the Ca$^{2+}$ binding pocket as observed in our results. Due to drastically structural transitions occurring upon Tyr-113 and Tyr-115 in the recognition and binding processes,[45] the moderate polarity and local dynamics should be crucial for the expelling of water and the binding of adenine part. For the Ca$^{2+}$ binding pocket, where some functional water molecules buried in it, its high polarity environment forms a relative rigid hydrogen-bond network, including Asp-21, Asp-40, Thr-41, Glu-

43. This rigid hydrogen-bond network is responsible for the hydrolysis of the phosphate backbone of DNA and RNA. Therefore for the $Ca^{2+}$ binding pocket, high polarity and relatively slow local dynamics at the bottom of this pocket are required to provide accurate and efficient nucleophilic attack.

## 6. Conclusion and outlooks

The picosecond time-resolved solvation dynamics at internal site and cavities of protein are reviewed here. We discussed the methods for studying the protein dynamics, with external and intrinsic probes. The external probe is versatile but cannot be site-specified. It may induce local environment variation for the detection. The intrinsic probe is feasible for site-specified study and can be used for mapping the whole protein molecule. However, because of the size and hydrophobicity of Trp residue, it may affect the overall protein integrity, which should be carefully handled before further studies. The solvation dynamics in the interior of SNase are successfully characterized using picosecond-resolved emission spectra of Trp as an intrinsic probe. The probes are buried at four internal sites. One is in the β-barrel, two sites close to the bottom of the nucleotide binding pocket, and one at the bottom of the $Ca^{2+}$ binding pocket. For all selected sites, two robust TRES processes of the intrinsic Trp residue on a few picoseconds and tens of picoseconds have been observed. The initial solvation relaxation in a few picoseconds is the fundamental relaxation of local water-network; the second solvation dynamics in tens of picoseconds represents the subsequent water-network rearrangements, coupled with local protein fluctuations, from a non-equilibrium configuration to an equilibrated state. Trp probes at these sites show distinct differences in solvation time scale and energy relaxation. Solvation relaxation at the bottom of the $Ca^{2+}$ binding pocket has the slowest dynamic time scales ($\tau_1 \sim 10$ ps, $\tau_2 \sim 104$ ps) accompanied with the largest energy changes. Solvation relaxation in the hydrophobic core (β-barrel) shows the fastest ones ($\tau_1 \sim 5$ ps, $\tau_2 \sim 80$ ps) accompanied with the least energy changes. These distinct differences in time scale and energy change of solvation relaxation reflect diverse and characteristic dynamic features of different regions, which closely relate to protein stability and the recognition, binding and hydrolysis of substrates. The method of site-directed mutagenesis combined with ultrafast spectroscopy is a unique way to map and fully understand the fundamental physics in protein.

## 7. Acknowledgement

The wild type (WT) SNase plasmid was generously provided by Prof. Bertrand Garcia-Moreno (Johns Hopkins University). We are very appreciate Dr. Lijuan Wang (Ohio State University) directing us prepare all the mutants and setup of biological lab. This work supported by the National Key Basic Research Program of China 2013CB921904, 2009CB930504, 2013CB328700; National Natural Science Foundation of China under grant Nos 11074016, 11121091, 10934001, 61177020，11134001, 10828407.